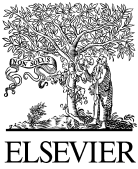

# Power Management in Smart Residential Building with Deep Learning Model for Occupancy Detection by Usage Pattern of Electric Appliances

Sangkeum Lee[a], Sarvar Hussain Nengroo[b], Hojun Jin[b], Yoonmee Doh[a], Chungho Lee[a], Taewook Heo[a], Dongsoo Har[b],*

[a]*Environment ICT Research Section, Electronics and Telecommunications Research Institute (ETRI), Daejeon 34129, Republic of Korea*
[b]*Cho Chun Shik Graduate School of Mobility, Korea Advanced Institute of Science and Technology (KAIST), Daejeon 34141, Republic of Korea*

**Abstract**

With the growth of smart building applications, occupancy information in residential buildings is becoming more and more significant. In the context of the smart buildings' paradigm, this kind of information is required for a wide range of purposes, including enhancing energy efficiency and occupant comfort. In this study, occupancy detection in residential building is implemented using deep learning based on technical information of electric appliances. To this end, a novel approach of occupancy detection for smart residential building system is proposed. The dataset of electric appliances, sensors, light, and HVAC, which is measured by smart metering system and is collected from 50 households, is used for simulations. To classify the occupancy among datasets, the support vector machine and autoencoder algorithm are used. Confusion matrix is utilized for accuracy, precision, recall, and F1 to demonstrate the comparative performance of the proposed method in occupancy detection. The proposed algorithm achieves occupancy detection using technical information of electric appliances by 95.7~98.4%. To validate occupancy detection data, principal component analysis and the t-distributed stochastic neighbor embedding (t-SNE) algorithm are employed. Power consumption with renewable energy system is reduced to 11.1~13.1% in smart buildings by using occupancy detection.

## 1. INTRODUCTION

Development of smart applications and services depends increasingly on the availability of occupancy statistics in residential buildings, offices, or educational facilities. Broad range of sensors which have been placed in recent years make it possible to obtain this kind of information. These sensors can gather various types of data, including

---

temperature, air quality, and even vibrations. With such data, it is possible to learn more about the residential buildings and others, and create fully intelligent applications.

The U.S. department of energy has published statistics showing that buildings use more than 70% of all the electricity produced annually [1]. The awareness of increasing energy efficiency in many areas, including lighting, information technology, heating, ventilation and air-conditioning (HVAC) systems, energy management in buildings, and so forth, has recently been in the spotlight [2, 3]. The effectiveness of HVAC systems has a significant impact on energy usage among the numerous features of buildings as a whole [4]. On the other hand, when actual occupancy differs from the planned schedule, certain buildings with schedule-based programmable thermostats may use more energy than those without smart apparatus. Different methodologies have been used for the development of automatic thermostat control systems while constructing more energy efficient systems [5].

Generally, the world's current energy consumption is an issue for both civilization and the environment because most of the energy is generated by combustion, which uses non-recyclable resources and emits unwanted greenhouse gases (GHGs) into the atmosphere [6]. To solve this problem and reduce its negative repercussions, energy efficiency and conservation are necessary. The development of alternative energy sources, such as wind turbine farms, tidal turbines, photovoltaic (PV) cells, or geothermal installations, presents another issue as it needs to consolidate the coexistence with the conventional energy resources, such as nuclear, oil, or coal. In economic terms, it is also essential to find ways to lower the costs relevant to power consumption.

Different components, such as plug loads from appliances, HVAC system, information technology devices, etc., must be taken into account when designing a smart building for residence or business or academic institute. Energy management systems, one of the key components of smart buildings, must improve energy efficiency and should be integrated with smart grid technology. Today's smart grid technologies, including microgrids [7], demand-side management (DSM) [8], load scheduling techniques, peer-to-peer (P2P) electricity trading [9], energy storage services [10], energy hubs, and energy prosumers, renewable energy resources (RES), complicate the functionality of the energy management system (EMS).

To operate building energy systems optimally and increase energy efficiency, occupancy information is crucial. Recent studies [2, 11] have anticipated that precise occupancy data in a building could result in energy savings in the range of 20% to 50%. Depending on the adaptive thermal comfort model [12], simulation results demonstrated that availability of occupancy information can lead to energy saving from 11% to 34% at various climate zones without causing additional user discomfort [13]. Through the inclusion of new features to better manage energy devices, the recent development of smart home products could be advantageous for the residential market. Lacking a reliable occupancy detection system, the present products could have the disadvantages of privacy issues, loss, or increased user discomfort. In [14], it is shown that only 12.5% of the U.S. residential homes employ smart home equipment such as thermostats, refrigerators, washers, and dryers due to the low return on investment.

With a flexible, effective, long-lasting and secure architecture, smart grid is the primary facilitator of innovative power grids. Optimal control of all the components that make up electricity systems is performed because of the new advancement in communication, sensing, and metering technologies that fall under the smart grid typology. Buildings are one of the power grid sectors that garner the most attention due to their high electricity consumption rate and potential for energy savings. Different studies have demonstrated the significance of occupancy information for enhancing building energy efficiency and lowering energy use [11, 15]. Agarwal et al. [16] proposed an HVAC control technique that turns the system ON or OFF according to occupancy detection in offices. The simulation results show that by employing the proposed method, the HVAC energy consumption is decreased by 10~15%. A control ventilation approach depending on occupancy information has the potential to save 55% of the energy used by the ventilation system [17]. By utilizing a network of passive infrared (PIR) and illumination sensors for occupancy detection, it is demonstrated in [18] that it is possible to save 3.5 hours of lighting electricity per a day in a campus room. The PIR and radio frequency identification sensors are used in [19] to forecast home occupancy patterns for heating control. Three distinct control methods based on always-on, scheduled, and PreHeat algorithms are examined and as a result energy savings of up to 35% is obtained. Different statistical classification models, primarily linear discriminant analysis, classification and regression trees, and random forest are used by the authors on the dataset containing light, temperature, humidity and $CO_2$ values [20]. The results demonstrated that by incorporating information about the time of day and week, occupancy detection accuracy is increased by 32% with high accuracies about 97% when only two predictors are used.

Although energy consumption optimization and reduction applications have received a lot of attention, there are other fields that can benefit from the potential availability of occupancy information for the various regions of a building. An autoencoder, made up of a graph convolutional network and a bidirectional long short-term memory (LSTM) network, is employed to the smart metering technique to identify outliers and missing values in the data [7]. Attempts devoted to the intelligent management and control in buildings in day-to-day operations, such as assessment and prediction of occupancy level for evacuation processes and decision-making in emergency circumstances, or identification of incidents for surveillance purposes have been reported in [21, 22]. Indoor comfort of resident is another area and for the comfort of resident occupancy information is crucial. High-resolution meter readings can be used to deduce the occupancy status of buildings, as a result of the recent installation of advanced metering infrastructures (AMIs) which monitor electrical power usage over time. The ability to infer occupancy from meter data in real-time is quite beneficial for managing appliances as it does not necessitate the installation of extra sensors. However, it also raises the possibility of burglary and exposes household working, eating, and vacationing routines in the event that the meter readings are tampered with [23]. As a result, it is critical to comprehend how accurately smart meter data can forecast whether or not a household is occupied. It would be encouraging for adaptive appliance control if meter data could be found to improve prediction accuracy of a building's occupancy status. But, it would also suggest the need for extra privacy protection measures for homeowners.

Models handling the time series data include deep learning models like LSTM networks and recurrent neural networks, as well as some widely used algorithms from the pre-deep learning era such as logistic regression and k-nearest neighbors. However, each of these approaches has some drawbacks. The LSTM network requires a lot of time and memory during training, like the majority of deep learning models. One of the main causes is that the model in recurrent layers sequentially receives inputs from subsequent timesteps to compute the layer weights, and the computation time grows with increase in number of input timesteps. In addition, this is susceptible to overfitting and vanishing gradient problems, making training and modification a challenging task. Apart from that, other algorithms like logistic regression and k-nearest neighbors fail to capture the temporal correlations in the input data.

This study uses machine learning to detect occupancy, based on the technical data of electric appliances in residential buildings. A novel method of occupancy detection for smart residential building system is proposed. The dataset, which is gathered from 50 residences and includes data obtained from sensors, light, HVAC, and electric appliances, is used in simulations. The autoencoder technique is used to categorize the occupancy among datasets provided from the smart metering systems. Confusion matrix is utilized for accuracy, precision, recall, and F1 to demonstrate the comparative performance of the autoencoder algorithm in occupancy detection. PCA and t-SNE technique are used to validate occupancy detection data.

The structure of this study is as follows. Section II presents the structure of the smart residential building system. The occupancy detection utilizing the autoencoder algorithm is presented in Section III. The simulation results of occupancy detection for smart residential building system are described in Section IV. In Section Ⅴ, concluding remarks are presented.

## 2. SMART RESIDENTIAL BUILDING SYSTEM

This section outlines the architecture that we have used to detect occupants in real time. As shown in Fig. 1, it is created to be as generic as possible to be used in various smart buildings that need real-time monitoring and processing.

Fig. 1(a) shows that the entire smart home is composed of sensors and electric appliances. The framework consists of sensing network, which measures temperature, humidity, $CO_2$, PM2.5, PM10, TVOC and illuminance, and measurement and controlling units. Energy gateway collects data measured from lighting, sensors, air conditioners, and electronics. Moreover, it controls activation/deactivation of home appliances, generates dashboard to give feedback about power usage, and provides a ubiquitous connection to the broadband internet. Through data collection, a robust smart home system is established over four-step measurement sensing, analysis group, optimization and recommendation, and control operation. Different types of sensors used in the smart home shown in Fig. 1(b) and Table 1 are as:

1) Smart light switch (STM-300W) is a general light switch having ON/OFF function.
2) Smart dimming controller (LCM-300W) is suitable for managing and controlling dimming lights.

3) Smart heating control (BCM-300W) can control heating from a smartphone for home or building. It requires a set up that connects to Wi-Fi network which allows to control the temperature of individual rooms wirelessly and remotely.

4) Smart outlet (CCM-300W) protects from overvoltage, over current, high power, etc. and has the appliance ON/OFF function

5) Smart power meter (PMM-300W) can monitor the current, voltage, power, frequency, and power factor in real time, and the measurement error is 0.5% of the power.

6) Smart air conditioner (ACM-300W) can perform ON/OFF function of air conditioner, temperature, set temperature, cold, heating and user remote control through IR coverage.

7) Home gateway (HGW-300) is a device that interlocks and manages sensors such as door sensors, motion sensors, smart buttons, and temperature/humidity sensors using Wi-Fi, ethernet, and ZigBee.

8) Smart air quality measuring instrument (AQM-300W) can comprehensively check and manage indoor air conditions by measuring temperature, humidity, fine dust (PM10), ultrafine dust (PM2.5), $CO_2$, gas chemicals (TVOC), and illumination. It connects to the smart home gateway through Wi-Fi communication.

9) Multifunctional Sensor (USM-300ZB) is a device that measures temperature, humidity, illumination, and movement.

10) Door Sensor (DSM-300ZB) is device that detects the opening/closing of door.

As the adoption of RESs in buildings increases, the installation of solar power is increasing. Among them, building-integrated renewable energy system is attracting worldwide attention as a way to increase energy efficiency and achieve reduction of GHG emissions. The Max, 75%, 50%, 25%, Min percentile of renewable energy production are shown in summer in Fig. 2. In summer, solar energy production is high, but depending on the cloud effect and rain, etc., partial shading patterns occur on the surface of the module, so it can be seen that the amount of PV power generation varies.

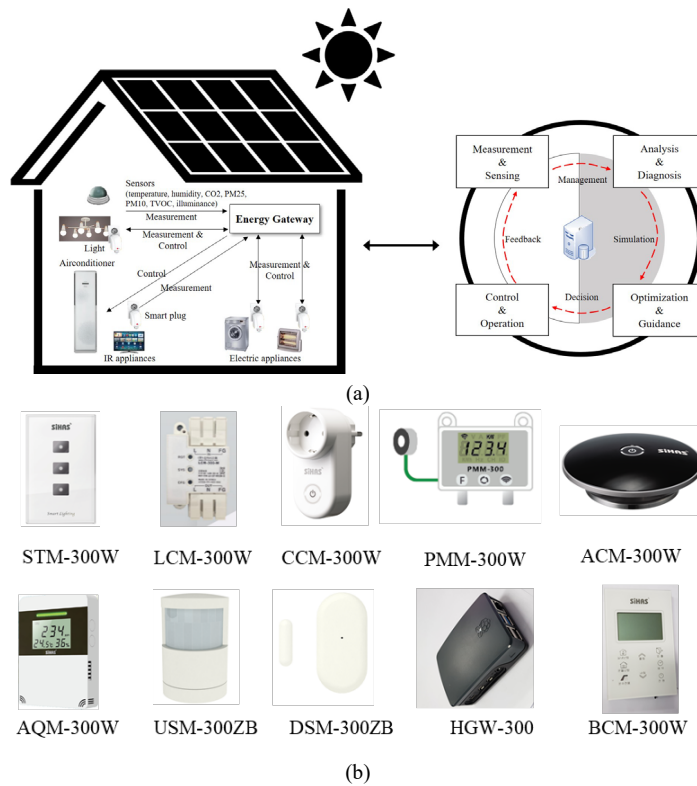

Fig. 1. Overall architecture of smart residential building system: (a) smart residential building system; (b) sensors in residential building.

Table 1. Sensors used in residential building system during the study.

| Type | Make | Model | Qty. |
|---|---|---|---|
| Light | Shin-A Systems | STM-300W | 4*50 |
| Dimming | Shin-A Systems | LCM-300W | 4*50 |
| Smart Heating Control | Shin-A Systems | BCM-300W | 1*50 |
| Smart outlet | Shin-A Systems | CCM-300W | (5~20)*50 |
| Smart Power Meter | Shin-A Systems | PMM-300W | 1*50 |
| HVAC | Shin-A Systems | ACM-300W | 1*50 |
| gateway | Shin-A Systems | HGW-300 | 1*50 |
| Smart air quality measuring instrument | Shin-A Systems | AQM-300W | 4*50 |
| Multifunctional Sensor | Shin-A Systems | USM-300ZB | 4*50 |
| Door Sensor | Shin-A Systems | DSM-300ZB | 4*50 |

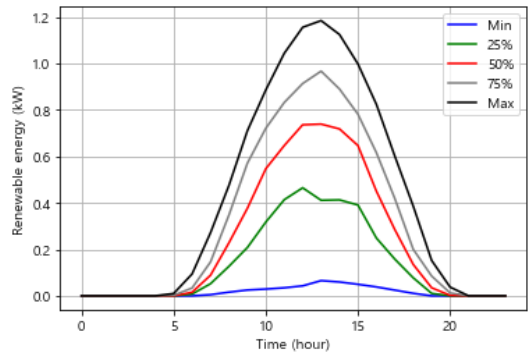

Fig. 2. Renewable energy production in summer.

## 3. OCCUPANCY DETECTION USING SVM AND AUTOENCODER

Support vector machines (SVMs) have been extensively used in classification, forecasting, and regression of random data sets [24]. Strong theoretical foundations based on Vapnik-Chervonenkis theory [25] are responsible for their practical success. The ability of SVM to employ a kernel function to map non-linear functions from a low-dimensional space to a higher-dimensional one is one of its key features. The fundamental goal of regression analysis is to find a function that can reliably predict future values. The general support vector regression (SVR) estimating function is given as

$$f(x) = w \cdot \Gamma(x) + \lambda$$

where $w \in \mathbb{R}^n$ is a hyperplane direction, $\lambda \in R, x \in \mathbb{R}^n$ is generated from the feature space, and $\Gamma$ stands for a nonlinear transformation from $R^n$ to a high dimensional. The ability to transfer a feature into a more complex dimension is provided by the transformation. By minimizing the regression risk, the value of $x$ can be solved as:

$$R_r(f) = C \sum_{i=0}^{l} G_i + \frac{1}{2} \parallel w \parallel^2$$

where $C$ is a constant and $l$ represents the size of training data. Loss function $G_i$ is given by:

$$G_i = \begin{cases} |f(x_i) - y_i| - \varepsilon, & |f(x_i) - y_i| \geq \varepsilon \\ 0, & otherwise \end{cases}$$

where, $x_i \in \mathbb{R}^n$ is the training data, $y_i$ is the target value and $\varepsilon$ is the insensitive loss.

In the smart residential buildings, occupancy detection can occur in complex patterns. An autoencoder is used to build the deep learning model and identify occupancy for power management. The autoencoder uses unsupervised learning, which obtain forecasting outcomes from incoming data by clustering data without labels of correct responses. An encoder and a decoder that both use the network to process the data sequence are combined into an autoencoder. The input sequence is sequentially entered to the encoder. The decoder regenerates the input sequence following the previous input sequence or produces a predicted value for the target sequence. Occupancy data are used for the autoencoder training, and the output is produced to calculate the reconstruction error by contrasting the output sequence data with the input sequence data. Autoencoders are trained to minimize the reconstruction error which can be also described as the squared error. The reconstruction error is presented as follows:

$$e_t = ||x_t - \hat{x}_t||^2$$

where $e_t$ is the reconstruction error, $x_t$ and $\hat{x}_t$ are vector elements of input and output (sensors, light, HVAC, and electric appliances)

In recent years, IoT technologies have been utilized to link many building sub-systems with additional environmental & contextual sensors, including CO2, temperature, humidity, and motion sensors. Using IoT technologies, it is possible to keep track and gather useful information of the building that might be used to detect occupancy [26, 27]. By establishing the HVAC and lighting system control strategies, real-time indoor occupancy information in buildings becomes a crucial role in enhancing occupant comfort and significantly reducing energy usage [28].

As presented in Fig. 3, CO2, humidity, illumination, PM10, PM25, temperature, TVOC, and presence can be confirmed by AQM and USM installed throughout the smart home in 50 households. It is possible to determine the presence or absence of the occupant through the information of the sensor. It is shown that CO2, PM10, and PM25 are high in the evening and in the morning, which represents the presence of occupant. It has a similar pattern which have a correlation with the occupancy.

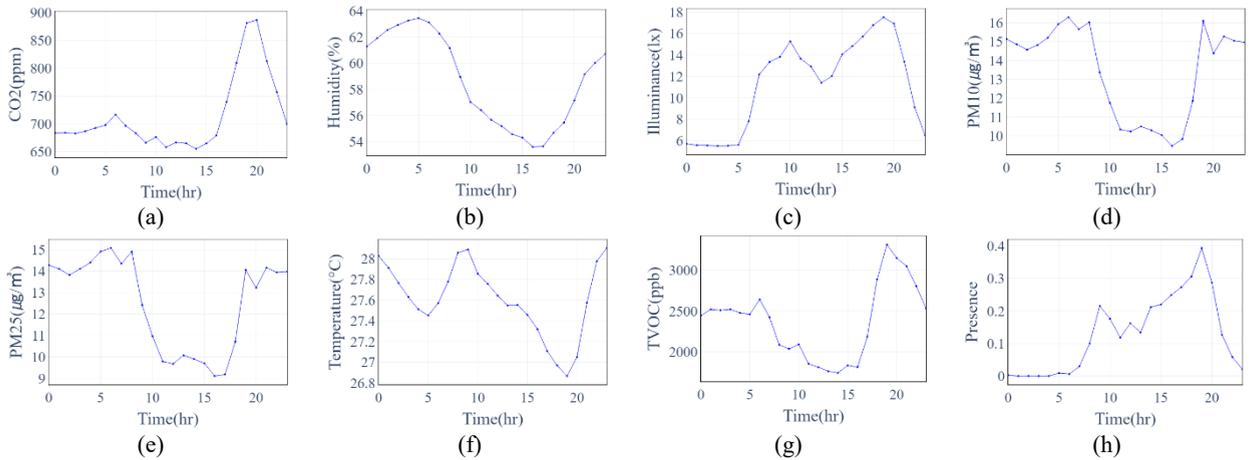

Fig. 3. Sensors dataset average days: (a) CO2; (b) Humidity; (c) Illuminance; (d) PM10; (e) PM25; (f) Temperature; (g) TVOC; (h) Presence.

As shown in Fig. 4, the average daily power and switch data patterns of PC, TV, washing machine, air purifier, electric rice cooker, microwave oven, coffee pot, and hair dryer can be observed. Looking at the power data of the PC and TV, it can be seen that they are actively used in the evening. With this correlation, it is used to predict occupancy.

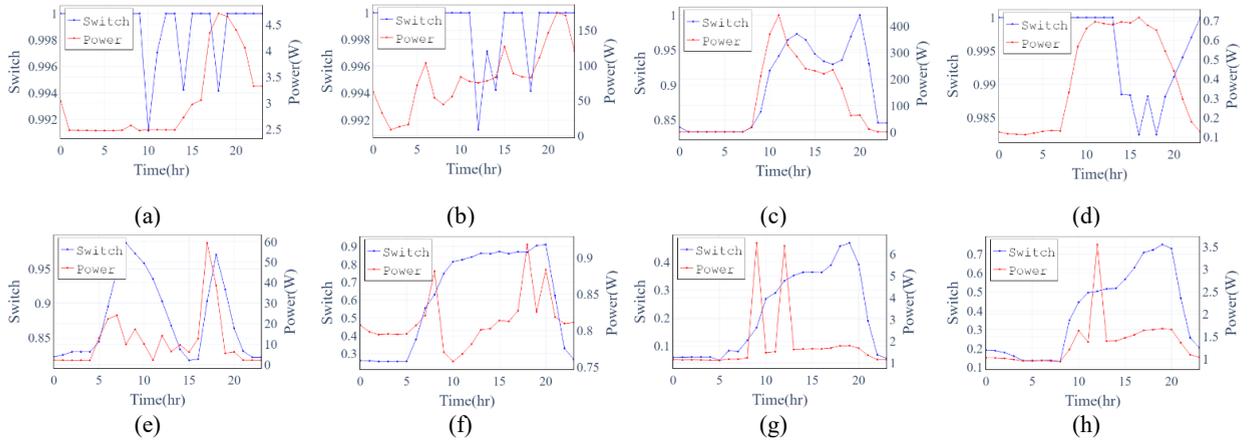

Fig. 4. Electric appliances average days: (a) PC; (b) TV; (c) Washing machine; (d) Air cleaner; (e) Cooker; (f) Microwave; (g) Coffee pot; (h) Hair-dryer.

## 4. SIMULATION RESULTS

This primary goal of this study is to evaluate and investigate building occupancy data by integrating different sensors and modern IoT technologies.

Table 2. Performance evaluation of SVM algorithm.

| Algorithm | Location | Accuracy | Precision | Recall | F1 |
|---|---|---|---|---|---|
| **SVM** | Room 1 | 0.984 | 0.887 | 0.885 | 0.886 |
| **SVM** | Room 2 | 0.983 | 0.954 | 0.950 | 0.952 |
| **SVM** | Room 1 & 2 | 0.957 | 0.903 | 0.901 | 0.902 |
| **Autoencoder** | Room 1 & 2 | 0.964 | 0.916 | 0.918 | 0.917 |

Temperature, humidity, $CO_2$, $PM_{25}$, $PM_{10}$, TVOC, illuminance, and HVAC, light, and electric appliance's smart plug switch are used as input of the SVM algorithm. To train SVM, 80% and the remaining 20% of data samples are used for training and validation, respectively. Accuracy, precision, recall, and F1 defining the confusion matrix are shown in Table 1. In confusion matrix, the terms are defined as: accuracy = (TP+TN)/(TP+FN+FP+TN), precision = TP/(TP+FP), recall = TP/(TP+FN), and F1 = 2*precision*recall/(precision+recall). In classification findings, the letters TP, TN, FP, and FN stand for true positive, true negative, false positive, and false negative, respectively.

As shown in Fig. 1, the proposed IoT and electronic product information-based occupancy detection method consists of three steps: (a) data collection, (2) data processing and storage, and (3) occupancy detection. As the occupancy sensor detects only simple movement information, a separate judgment is required for the occupant's space occupancy. In this study, the occupancy status is maintained according to the time range from the information of the occupancy sensor installed in the space, and the time range is divided into the active time zone (6 am to 24 pm) and the inactive time zone (0 am to 6 am). Resampling of electronic product usage information in the time zone is performed after every 15-minute intervals.

Based on the determined occupancy state, the switch on/off state of the home appliance being measured in the corresponding space is analyzed. It is possible to reduce the wasted electrical energy by operating the home appliance in the occupancy state and turning off the power to the home appliance in the absence state. Fig. 5 shows the occupancy detection algorithm according to the room location. Room 1 and room 2 are spaces of occupancy detection in living room and bedroom, respectively, in Fig. 5 and Table 2. As a result, using the data of room 1 and room 2, accuracies of 0.984 and 0.983 are obtained by SVM, respectively, when judging occupancy in Table 2. Accuracies of SVM and autoencoder algorithm for room 1 & 2 are obtained by 0.957 and 0.964, respectively.

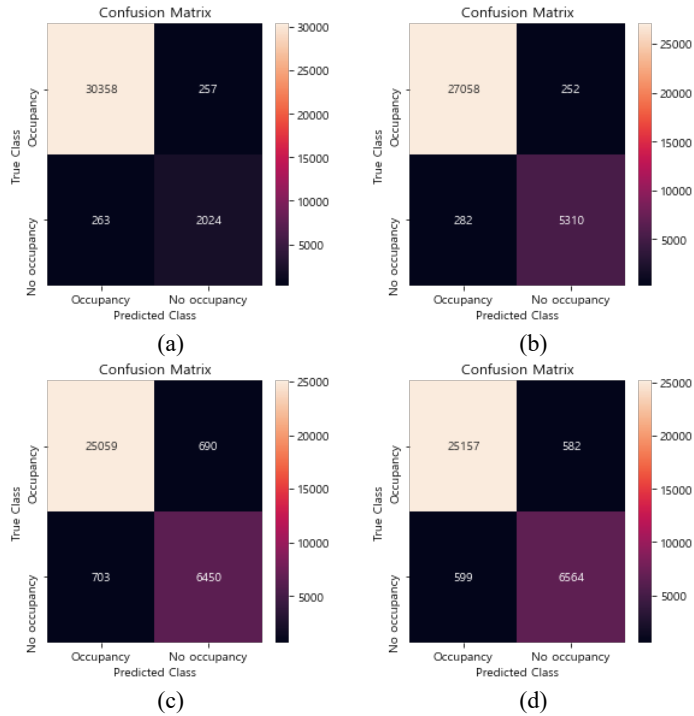

Fig. 5. Occupancy detection's confusion matrix in smart residential building system:
(a) Room 1 by SVM; (b) Room 2 by SVM; (c) Room 1 & 2 by SVM; (d) Room 1 & 2 by autoencoder.

When validating occupancy detection data, column standardization of the input variables is used to remove the influence of various variable dimensions. The dimensionality reduction operation is then visualized using the t-SNE technique, and the default output dimension is 2.

The main goal of the t-SNE algorithm is to minimize the KL divergence between the original space and the embedded space sample distribution. As the KL divergence is not convex, multiple iterations are necessary to reach the ultimate stable (converging) optimal solutions. It can be observed that the error is low enough to converge after 1000 iterations and produce stable and trustworthy result. The output findings using the PCA and t-SNE algorithms (both of which have two-dimensional target dimensions), after dimensionality reduction, can be seen in Fig. 6.

In Fig. 6(a), the red dot and the blue dot represent the data sample for "No occupancy" and "Occupancy", respectively. It is shown that the hyperplane of nonlinear support vectors can be used to differentiate the two, as many input variables are compressed into two-dimensional space through PCA and t-SNE algorithms and samples of different categories appear as blocks. PCA dimensionality reduction does not produce the best graphical result. There is a lot of sample overlap, making it challenging to identify the clusters close to the center. This might be as a result of the fact that the input variables for social and customer qualities are complicated nonlinear non-Gaussian in nature. The t-SNE approach, on the other hand, is more robust and uses a long-tailed t distribution to fit the distribution of data in low-dimensional space, better capturing the overall properties of the input data.

Based on this realistic situation, this model is first developed through the t-SNE algorithm. The retention of original data information is maximized while reducing many input properties affecting occupancy to two dimensions. It is easy to visualize, decreasing the complexity of the input variables significantly, and then utilizing a nonlinear SVM to train and forecast a reduced-dimensional sample. The empirical analysis of the occupancy reveals that the generalization and learning capabilities of the prediction model suggested in this study are largely acceptable, and it achieves accurate occupancy detection in a smart home system.

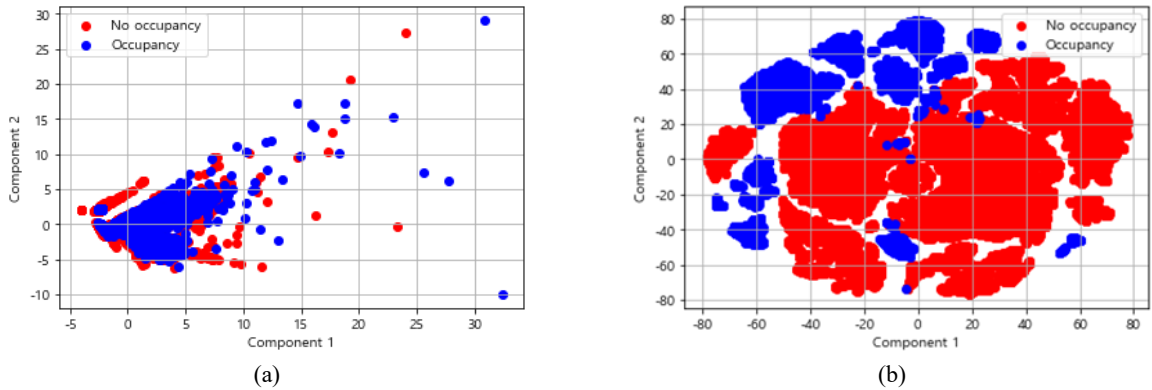

(a)　　(b)

Fig. 6. Two-dimensional display of raw data by: (1) PCA algorithm; (b) t-SNE algorithm.

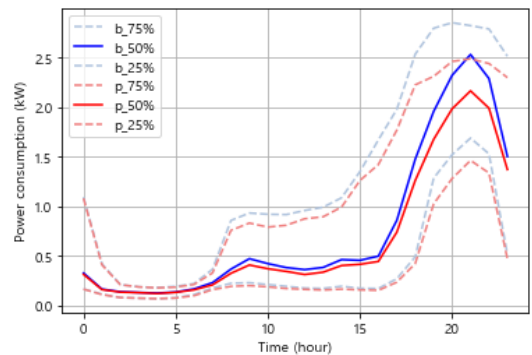

Fig. 7. Power consumption with renewable energy system in power management of smart buildings using occupancy detection.

In Fig. 7, 'b' is baseline without occupancy detection and 'p' is proposed algorithm with occupancy detection using autoencoder. The proposed algorithm detects occupants and turns off electronic products to perform power management. The 75%, 50%, 25% household percentile of total baseline power consumption is 30.82W, 18.15 kW, 9.913 kW, respectively. The 75%, 50%, 25% percentile total power consumption of the proposed algorithm is 27.40 kW (saving 11.1%), 15.89 kW (saving 12.4%), 8.61 kW (saving 13.1%), respectively. The highest saving effect is shown in the 25% percentile of power consumption. It can be seen that a house that uses a lot of electricity has a small amount of power reduction because it uses a lot of electricity and continues to use home appliances.

## 5. CONCLUSION

This study proposes a novel approach to occupancy detection for smart residential building systems. The dataset, which is collected from 50 residences and consists of data obtained from electrical appliances, sensors, lighting, and HVAC, is used in simulations. The autoencoder technique is used to categorize the occupancy among datasets measured by smart metering systems. To prove the performance of the SVM and autoencoder algorithm in occupancy detection, confusion matrix is used for accuracy, precision, recall, and F1. The SVM and autoencoder algorithm achieves performance of occupancy detection using technical information of electric appliances by 95.7~98.4%. For validation of occupancy detection data, PCA and the t-SNE algorithm are used. In future work, we plan to derive an advanced occupancy detection technique that integrates information from various rooms. Power consumption with renewable energy system is reduced in smart buildings using occupancy detection by 11.1~13.1%.

## Acknowledgements

This work was supported by the Korea Institute of Energy Technology Evaluation and Planning (KETEP) and the Ministry of Trade, Industry & Energy (MOTIE) of the Republic of Korea (No. 20192010107290). This work was also supported by the Energy Cloud Research and Development Program through the National Research Foundation of Korea (NRF) funded by the Ministry of Science, ICT, under Grant 2019M3F2A1073314.